\documentclass[aps,prl,reprint,superscriptaddress]{revtex4-2}

\usepackage{amsmath}
\usepackage{siunitx}
\usepackage[inline]{enumitem}

\usepackage{xcolor}  
\usepackage{graphicx}

\graphicspath{{Images/}}

\begin{document}


\title{Photonic Topological Baths for Quantum Simulation}


\author{Abhi Saxena}
\author{Yueyang Chen}
\author{Zhuoran Fang}
\affiliation{Department of Electrical \& Computer Engineering, University of Washington; Seattle, Washington, 98195, USA}
\author{Arka Majumdar}
\email[Corresponding author: ]{arka@uw.edu}
\affiliation{Department of Electrical \& Computer Engineering, University of Washington; Seattle, Washington, 98195, USA}
\affiliation{Department of Physics, University of Washington; Seattle, Washington, 98195, USA}


\date{\today}

\begin{abstract}
Quantum simulation involves engineering devices to implement different Hamiltonians and measuring their quantized spectra to study quantum many-body systems. Recent developments in topological photonics have shown the possibility of studying novel quantum phenomena by controlling the topological properties of such devices. Here, using coupled arrays of upto $16$ high Q nano-cavities we experimentally realize quantum photonic baths which are analogs of the Su-Schrieffer-Heeger model. We investigate the effect of fabrication induced disorder on these baths by probing individual super-modes and demonstrate the design mitigation steps required to overcome the disorder effects on the quantum phenomena.
\end{abstract}


\maketitle

\begin{figure}[h]
	\includegraphics[width=1\linewidth]{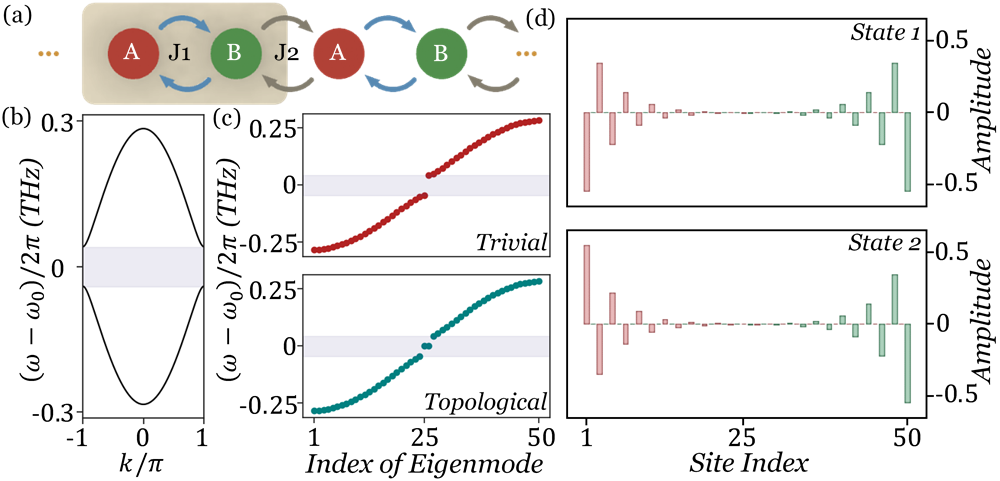}
	\setlength{\abovecaptionskip}{-5pt}
	\setlength{\belowcaptionskip}{-5pt}
	\caption{\label{fig:fig1001} (a) Schematic of the SSH bath. The photonic lattice consists of two sub-lattices $A$ (red) and $B$ (green). The intracell hopping rate is $J_1$ and intercell hopping rate is $J_2$. (b) Dispersion relation of the SSH bath where the shaded region denotes the band gap. (c) Resonant frequencies of a finite bath consisting of $N=50$ sites. Trivial phase (red) is characterized by $J_1>J_2$, whereas the topological phase (blue) characterized by $J_1<J_2$. (d) Wave function of the hybridized edge modes of a topological bath with $N=50$ sites. Left edge state is localized on sub-lattice $A$ (red), right edge state is localized on sub-lattice $B$ (green).}
\end{figure}

Use of photons as particles in controlled quantum systems to study other complex non-equilibrium quantum phenomenon forms the basis of one of the most promising paradigms for quantum simulation \cite{georgescu2014quantum}. This typically involves implementing Hamiltonians by engineering coupled photonic resonators to tailor photon’s energy momentum relationships \cite{roushan2017spectroscopic,kollar2019hyperbolic}. Recent advances in controlling topological properties of photonic lattices have shown that novel forms of light-matter interaction can be realized in such systems owing to the topological protection of the resulting quantum many body states \cite{bello2019unconventional,anderson2016engineering}. While the microwave photons in superconducting circuits have already been used to exhibit these unconventional quantum phenomena \cite{kim2021quantum}, such a demonstration is missing in optical domain. Observing similar effects in optics will not only significantly simplify the experiments due to performability at much higher temperatures but will also allow measurements of multi-particle correlations due to the availability of single photon detectors. In fact, recent demonstrations of large-scale photonic quantum computers attest to such inherent scalability of photonics \cite{arrazola2021quantum,zhong2020quantum}.
\\ 
To reach this regime of interacting photons for quantum simulation in optical domain we need to implement the topological phases on photonic lattices made with high quality (Q) factor and low mode volume resonators with spectral accessibility to individual super modes \cite{hartmann2016quantum}. However, due to unavoidable imperfections in nanofabrication, high Q optical resonator arrays are inherently prone to disorder in their resonant frequencies. Without undertaking suitable mitigating steps in design of these arrays, this uncontrolled disorder can be a huge impediment in constructing topological baths suitable for studying various quantum optical phenomena. Typical coupled cavity arrays used to demonstrate topological states in the optical domain \cite{mittal2019photonic,mittalPhdthesis,parto2018edge,st2017lasing} operate in a regime where fabrication induced disorder is comparable or greater than the relevant hopping rates between the cavities. In this paper, by increasing the effective mode overlap between resonators sites we overcame the effects of the underlying disorder and experimentally realized topological quantum electrodynamic baths which are photonic analogs of the Su-Schrieffer-Heeger (SSH) model \cite{su1979solitons}.  We show that these coupled cavity array baths (with individual Q-factor exceeding $3.1\times{10}^4$) operate in a regime which is suitable for quantum simulation and can be used to impart special topological properties to interacting photons as discussed in depth by \citet{bello2019unconventional}. 
\\
The SSH model describing the topological photonic bath is illustrated in Fig. \ref{fig:fig1001}a. The photonic lattice consists of sub-lattices $A$ and $B$ of the array respectively, made up of cavities with resonant frequency $\omega_0$. The intracell hopping rate between the sites $A$ and $B$ of a unit cell is given by $J_1$ and the intercell hopping rate between the unit cells is denoted by $J_2$. The Hamiltonian of this bath can be written as ($\hbar=1$)
\begin{align}\label{hamiltonian}
	\mathcal{H}_B =\sum_{i}  \omega_0 (a_i^\dag a_i + b_i^\dag b_i) + J_1(b_i^\dag a_i + a_i^\dag b_i) \nonumber \\ + J_2 (b_i^\dag a_{i+1} +a_{i+1}^\dag b_i) 
\end{align}
where $ a_i^\dag(a_i)$ and $b_i^\dag(b_i)$ denote the site bosonic creation (destruction) operators at site $A$ and $B$ of the $i^{th}$ unit cell. Assuming periodic boundary conditions the Hamiltonian in momentum space can be written as  
\begin{align}\label{k_ham}
	\tilde{\mathcal{H}}_B(k)=\begin{bmatrix} \omega_0 & J_1 + J_2 e^{-jk} \\ J_1 + J_2 e^{jk} & \omega_0 \end{bmatrix}
\end{align}
The properties of this disorder free bath can be summarized as follows:
\renewcommand{\labelenumi}{(\roman{enumi})}
\begin{enumerate}[wide, labelindent=0pt, noitemsep, topsep=0pt]
	\item The bath has a chiral symmetry \cite{asboth2016short} owing to which the eigenstates of the Hamiltonian $\tilde{\mathcal{H}}_B(k)$ form two symmetric bands about $\omega_0$ given by  \[\omega_\pm(k)=\omega_0\pm\sqrt{J_1^2+J_2^2+2J_1J_2cos(k)}\]
where the $\pm$ denotes the upper/lower pass bands (Fig. \ref{fig:fig1001}b) with a band gap of $2|J_1-J_2|$.
	\item This bath supports topologically non-trivial phases \cite{asboth2016short} depending on whether $J_1<J_2$ which is termed as the ‘topological’ phase or $J_1>J_2$ which is termed as the ‘trivial’ phase. In a finite bath, these phases lead to formation of topological states localized at the edges of the lattice with an exponential decay into the bulk \cite{parto2018edge,asboth2016short}. These edge states lie in the middle of the band gap centered around $\omega_0$ for the topological phase and disappear for the trivial phase (Fig. \ref{fig:fig1001}c). They have non-vanishing amplitude over only one of the two sub-lattices $A$ or $B$ and the small amount of overlap in the bulk causes these edge modes to hybridize as even and odd eigenstates of the system (Fig. \ref{fig:fig1001}d).
	\item \citet{bello2019unconventional} and \citet{kim2021quantum} have shown that if a quantum emitter with transition frequency lying at the center of band gap ($\omega_0$) is coupled to a site of this SSH bath, the resulting photonic bound state mimics an effective topological edge state in middle of the bath. This bound state inherits all the properties of the topological edge states, localizing the photon to one direction and sub-lattice depending on the site type ($A$ or $B$) to which the emitter is coupled. In presence of several emitters these states can be used to mediate directional, topological interactions between them which can give rise to exotic many body phases.
\end{enumerate}

These properties, which are critical to realize unconventional quantum phenomena using the topological bath are adversely affected in the presence of fabrication disorder. In optical domain, a scalable implementation of the SSH bath relies on nanofabrication to create solid state optical cavity arrays. The major form of disorder that exists in solid state photonic systems is in the resonance frequencies of the cavities arising from the inconsistencies in lithography and etching as well as from the non-uniformities in the film thickness of the bare wafer itself. This disorder is modelled by modifying the diagonal terms of the Hamiltonian to include random $\delta_i$' s drawn from a Gaussian distribution with a standard deviation $\sigma$ \cite{underwood2012low,majumdar2012design}. Thus, the modified bath's Hamiltonian in presence of this diagonal disorder can be written as
\begin{align}\label{diag_dis}
	\mathcal{H}_B^{\sigma} = \mathcal{H}_B + \sum_{i} \Big (\delta_{a,i} a_i^\dag a_i + \delta_{b,i} b_i^\dag b_i \Big )
\end{align}
It is straightforward to see that the eigen-properties of this Hamiltonian depend on relative values of $J_1-J_2$, $\sigma$ and $J_1+J_2$. Hence, we define a dimensionless parameter
\begin{align}\label{dimen}
	\eta = 2\sigma/(J_1+J_2)
\end{align}
as the measure of relative disorder present in our system. We study the effect of the diagonal disorder on the key properties of the bath by looking at the evolution of topological bound states and the transmission spectrum as we sweep across $\sigma$ keeping $J_1$, $J_2$ constant.\\
As mentioned in property (iii), the presence of a quantum emitter lying in the middle of the band gap ($\omega_0$) induces formation of bound directional states which are akin to topological edge states in their properties \cite{bello2019unconventional}. We first study the effect of disorder on this type of bound state. 
\begin{figure}[b]
	\vspace{-10pt}
	\includegraphics[width=0.9\linewidth]{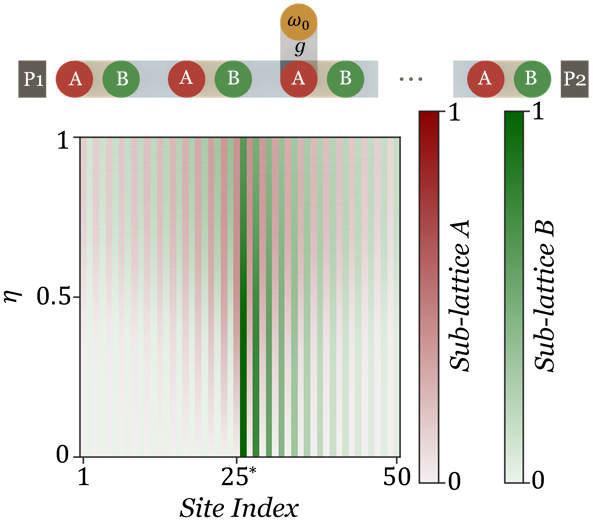}
	\setlength{\abovecaptionskip}{0pt}
	\setlength{\belowcaptionskip}{-15pt}
	\caption{\label{fig:fig1002} Modulus of amplitude of bound state wavefunction (normalized) as we sweep across $\eta$ averaging over ${10}^4$ disorder realizations per $\eta$. The schematic depicts a quantum emitter coupled to a SSH bath in trivial phase at a site of sub-lattice $A$. The coherent emitter-cavity coupling rate is $g=(J_1+J_2)/10$.  In absence of any disorder, the bound state is localized towards the right with non-zero amplitude on sub-lattice $B$ only. The direction of this envelope changes if the emitter is coupled to sub-lattice $B$ instead of $A$. ${25}^\ast$ denotes the position of emitter in the array. Amplitudes on sub-lattice $A$ are in red and on sub-lattice $B$ are in green.}
\end{figure}
\\
In Fig. \ref{fig:fig1002}, we consider a bath made of $50$ resonators configured in trivial phase with a quantum emitter coupled to a type $A$ site in the middle of the array. For this system, we plot the modulus of the amplitude of the induced bound state wave function averaged across ${10}^4$ disorder realizations per $\eta$. Without any disorder, the bound state envelope extends towards the right direction and occupies only the $B$ sub-lattice. Presence of diagonal disorder breaks the chiral symmetry of the Hamiltonian and protection to the bound state weakens. Despite this in the region where $\eta<<1$ the bound exists with strong localization towards one direction and vanishing amplitudes on the conjugate sub-lattice.  As the value $\eta$ increases, the bound state becomes more delocalized and has weights over both the sub-lattices. When $\eta\rightarrow1$ overall averaged bound state wave function over different realizations is spread across the array and loses all of its topological properties.
\\
\begin{figure}[h]
	\vspace{-10pt}
	\includegraphics[width=1\linewidth]{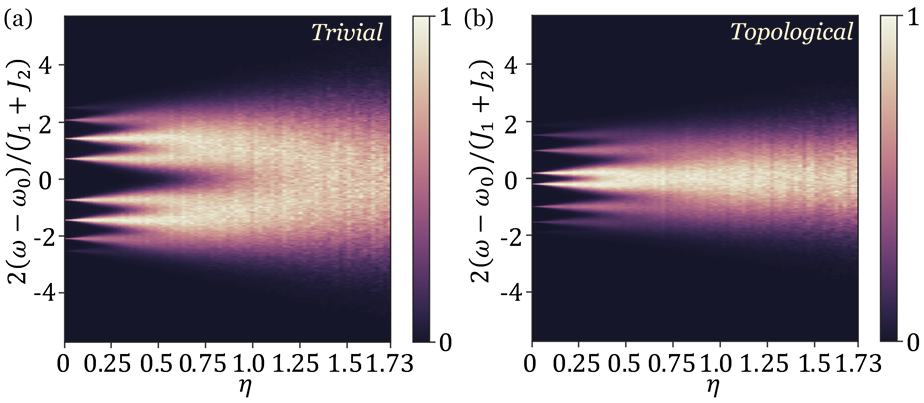}
	\setlength{\abovecaptionskip}{-10pt}
	\setlength{\belowcaptionskip}{-10pt}
	\caption{\label{fig:fig1003} Transmission spectrum $|S_{21}|^2$ of a photonic bath with $8$ sites; averaged across ${10}^4$ disorder realizations per $\eta$ (a) Trivial phase (b) Topological phase.  }
\end{figure}

Fabricated photonic baths are experimentally characterized by measuring the transmission spectrum $|S_{21}|^2$ as relevant bath parameters can be extracted in a straightforward manner from such a measurement. To observe the effect of disorder on the transmission spectrum; we simulate and plot $|S_{21}|^2$ of a photonic bath with $8$ sites for both topological and trivial phase, again averaging across ${10}^4$ disorder realizations per $\eta$ (Fig. \ref{fig:fig1003}). First thing that becomes immediately obvious is that the transmission amplitude of the farthest modes from the bare resonance is rather small even when $\eta=0$. This comes out naturally in the process of solving a set of coupled mode equations and combining it with the input-output formalism to express output field in terms of input fields \cite{Supp}. When the disorder is added to this model, we find that the transmission of the farthest modes gets rapidly less prominent even in the region $\eta<<1$. Consequently, probing all super modes becomes very difficult in a cavity array with large number of sites. As the value of $\eta$ further increases the modal peaks start to merge, and in the region with $\eta>>1$ the averaged plot approaches a broad Gaussian distribution. This behavior clearly indicates that $\eta<\ 1$ is a necessary condition when using the transmission spectrum for calculating the band gap for the trivial phase or identifying the edge modes from super-modes for the topological phase, both of which play critical role in determining quantum properties of the system.\\
Thus, looking at behavior of both the topological bound states and the transmission spectrum with increasing disorder, we can conclude that operating in the regime $\eta<<1$ can mitigate the effects of inherent fabrication disorder and allows us to harness the unconventional topological properties of the SSH bath.  This can be achieved by designing the bath such that individual sites have much larger hopping rates than the disorder.

\begin{figure*}[t]
	\includegraphics[width=1\linewidth]{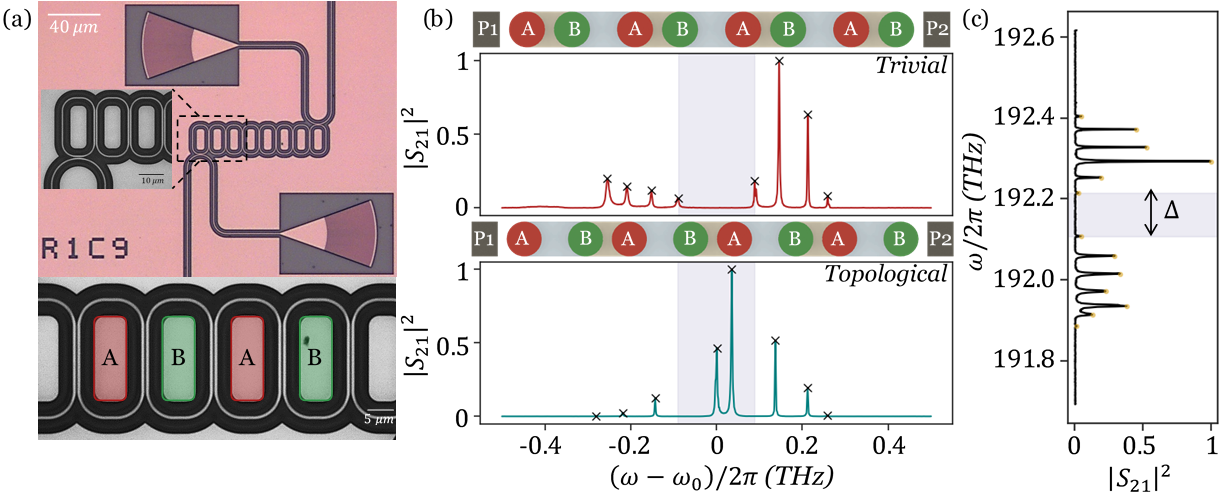}
	\setlength{\abovecaptionskip}{-10pt}
	\setlength{\belowcaptionskip}{-15pt}
	\caption{\label{fig:fig1004} (a) Optical micrographs of a photonic SSH bath. (b) Normalized transmission spectrum of a SSH photonic bath with $8$ sites. Upper plot (red) depicts the trivial phase, whereas the lower plot (blue) depicts the topological phase. We can observe the band gap (shaded) formation in trivial phase, and the existence of edge modes in middle of the band gap in topological phase. The schematic depicts the ports i.e., the grating couplers used for transmission measurement. (c) A SSH photonic bath made of $8$ unit cells i.e., $16$ sites in trivial phase. Measured band gap is $\Delta=\ 0.107\ THz$ which closely matches the theoretical prediction of $0.11\ THz$.}
\end{figure*}

To realize these baths, we implemented the photonic analog of SSH model as coupled cavity arrays made from racetrack resonators fabricated in silicon photonics with $220\ nm$ thick Si layer on a $3\ \mu m$ thick $SiO_2$ film (Fig. \ref{fig:fig1004}a). Each device is probed via a set of grating couplers located at the first and last site to coherently measure its transmission and reflection spectra. A similar platform has been used before \cite{mittal2019photonic} to show the existence of higher order edge states in photonic lattices. Those devices were operated in regime where localization lengths of the edge states were extremely small ($\frac{J_2}{J_1}>>1$) and the band gap was larger than the disorder. But their disorder to hopping rates ratio was  $\eta=1.95$ which is $>>1$. While edge states can be observed in this regime, as discussed above it is not suitable for enabling bath mediated directional interactions between emitters for quantum simulation purposes. In order to reach the regime of $\eta<<1$ we need to increase the absolute value of hopping rates between the lattice sites. Hopping rates between optical cavities depend on the mode overlap of the two resonators $J\propto e_1^\ast e_2$ \cite{haus1991coupled,smith2020active} where $e_1$ and $e_2$ denote mode volume normalized field profiles of the resonator modes. This overlap can be increased in broadly three ways:
\begin{enumerate*}
	\item by reducing the physical distance between sites,
	\item by increasing the length of coupling region between the resonators, and
	\item by reducing their mode volumes.
\end{enumerate*}
Resolution of the lithography process limits the closest gap we can place two resonators without introducing additional disorder. Therefore, to increase the hopping rates further we need to reduce the size of resonators and increase the length of interaction region between them.\\
To this end, we fabricated device arrays made up of racetrack resonators which are $60\ \mu m$ long with a $12\ \mu m$ long coupling region. To obtain two differing hopping rates between the resonators, we kept the inter-resonator gap to be either $90\ nm$ or $110\ nm$. The shorter length of the racetrack resonators not only helps in reduction of the mode volume but also ensures a large free spectral range for the devices to prevent interaction between different longitudinal modes of the same resonator.
\begingroup
\squeezetable
\vspace{-10pt}
\begin{table}[h]
	\caption{\label{table:tab1001}Disorder compilation of the SSH photonic bath}
	\renewcommand{\arraystretch}{1.5}
	\begin{ruledtabular}
		\begin{tabular}{cccccc}
			$\sigma_{type}$ & $J_1/2\pi(GHz)$ & $J_2/2\pi(GHz)$ & $\sigma/2\pi(GHz)$ & $\eta$ & $\#$ Measured\\
			\hline
			Global & 163 & 122 & 70.68 & 0.49 & 25\\
			\hline
			Local & 163 & 122 & 8.02 & 0.056 & 18\\
		\end{tabular}
	\end{ruledtabular}
\vspace{-5pt}
\end{table}
\endgroup

Then, we measured the transmission spectrum of these devices. It is well-known that the devices fabricated on different parts of the chip are more susceptible to disorder and suffer from an overall mean frequency shift owing to variations in nanofabrication processes across the chip area \cite{mittalPhdthesis,underwood2012low,majumdar2012design}. We label this as the global disorder and characterized it via a statistical study of spectral modes across devices and tracking their mean frequency (Table \ref{table:tab1001}). We fabricated arrays with varying number of racetrack resonator sites ranging from $1$ to $16$ on the same chip. Only devices in which we could spectrally resolve all the modes, were used to calculate the disorder. Usually this global disorder, calculated from statistics of the mean frequency of modes is subtracted from the spectrum as an overall shift to the origin \cite{underwood2012low}. This approach is strictly valid only if we are dealing with arrays that have small footprint on a chip. As the cavity arrays have an increasing number of sites, the device area starts to grow, and therefore within one big device global disorder cannot just be ignored as an overall origin shift. 
\\
We also calculated and characterized local disorder which persists even after shifting the spectra to a common origin. To calculate this disorder, we measured the deviation of each individual mode across instances of similar devices across the chip:  
\begin{align}\label{local_disord}
\sigma_{local} = \frac{1}{2\pi}\sqrt{\sum\frac{(\omega_i-\omega_i^{mean}-\omega_{shift}^g)^2}{n}}
\end{align}
where $\omega_i$ is the frequency of the $i^{th}$ super-mode,  $\omega_{shift}^g$ is the global frequency shift to align the origins and $\omega_i^{mean}$ is the mean position of the super-mode across the same device design made at different locations of the chip. This method allowed us to combine statistical data for disorder from devices with various number of sites and with varying phases (topological/trivial). The calculated value of $\eta$ for local disorder was $0.056$ ($<<1$). 
\\
The larger hopping rates present in these arrays allow for reduction of the effect of disorder. In lattices with up to $8$ sites we observe spectral accessibility to all the modes and global disorder has minimal effects on devices. This allows for clear comparison of band gap formation in the case of trivial phase and existence of two edge modes lying inside the band gap in the topological phase (Fig. \ref{fig:fig1004}b). For photonic baths with more than $8$ sites all modes are not clearly observable due to vanishing amplitudes further away from the mean frequency and increasing effects of global disorder on the spectrum. Despite this, the device design allows for realization of photonic baths (here, trivial phase) with upto $16$ sites (Fig. \ref{fig:fig1004}c). We can observe a clear band gap formation with $\Delta=\ 0.107\ THz$ which is within $3\%$ of the theoretical prediction. As discussed before such a photonic bath can be used to endow several special properties to coupled emitters also described in detail by \citet{bello2019unconventional}.
\\
In conclusion, using coupled cavity arrays we experimentally demonstrated topological photonic baths which are optical analogs of the SSH model. We studied the effect of fabrication induced disorder on these baths and demonstrated the steps required to overcome its effects. A similar, more detailed demonstration has been reported in superconducting systems \cite{kim2021quantum} and our work enables a way to bring such topological baths to optics by developing a paradigm to harness recent advancements made in topological photonics and applying them to quantum simulation in optical domain. Future work will include independent tuning of each resonator site in the array and integration of quantum emitters \cite{chen2018deterministic} with the photonic baths to probe topological quantum phenomena.

\vspace{-5pt}
\section*{Acknowledgments}
\vspace{-10pt}
\begin{acknowledgments}
The research was supported by National Science Foundation grant NSF-QII-TAQS-1936100 and National Science Foundation grant NSF-1845009. Part of this work was conducted at the Washington Nanofabrication Facility / Molecular Analysis Facility, a National Nanotechnology Coordinated Infrastructure (NNCI) site at the University of Washington, which is supported in part by funds from the National Science Foundation (awards NNCI-1542101, 1337840 and 0335765), the National Institutes of Health, the Molecular Engineering \& Sciences Institute, the Clean Energy Institute, the Washington Research Foundation, the M. J. Murdock Charitable Trust, Altatech, ClassOne Technology, GCE Market, Google and SPTS.
\end{acknowledgments}

%
\end{document}



\title{Supplementary Material: Photonic Topological Baths for Quantum Simulation}

\date{\today}

\maketitle


\section{Calculating transmission spectrum of photonic baths}

To model the transmission spectrum of a coupled cavity array we use input-output formalism to express the outfield fields in terms of the input fields and combine it with a set of coupled mode theory differential equations which contain the system information. We calculate the transmission spectrum at a particular frequency by solving the following matrix equation:
\begin{widetext}
\begin{align}\label{system}
\begin{pmatrix}j\delta_1-j\Delta-\frac{\kappa}{2}-\frac{\gamma}{2}&jJ_1&0&0&\ldots\\jJ_1&j\delta_2-j\Delta-\frac{\gamma}{2}&jJ_2&0&\ldots\\\vdots&\ddots&\ddots&\ddots&\vdots\\\ldots&\ldots&0&jJ_1&j\delta_N-j\Delta-\frac{\gamma}{2}-\frac{\kappa}{2}\\\end{pmatrix} \begin{pmatrix}c_1\\c_2\\\vdots\\c_N\\\end{pmatrix}=\begin{pmatrix}-j\sqrt\kappa\\0\\\vdots\\0\\\end{pmatrix}
\end{align}
\end{widetext}
where $\delta_i$s’ denote the disorder in frequency at the $i^{th}$ site, $\mathrm{\Delta}=\omega-\omega_0$ denotes the relative frequency of the resonators with respect to input laser, $\kappa$ denotes the coupling rate to the grating couplers, $\gamma$ denotes the loss rate at each site, $J_1$ $\&$ $J_2$ denote the hopping rates between the resonators and $c_i$s’ denote the field amplitude at $i^{th}$ lattice site. The light is inputted into the system via grating coupler at the first site. The output power measured at the grating coupler at the last site is thus given by:
\begin{align}\label{transmission}
	|S_{12}|^2=|j\sqrt\kappa c_N|^2
\end{align}
This system of equations is solved ${10}^4$ times for each Gaussian standard deviation $\sigma$ from which $\delta_i$s’ are drawn. The collected output spectra are then averaged and normalized for the disorder plot in the paper (Fig. 3).

\clearpage
 
\begin{figure*}[t]
	\includegraphics[width=1\linewidth]{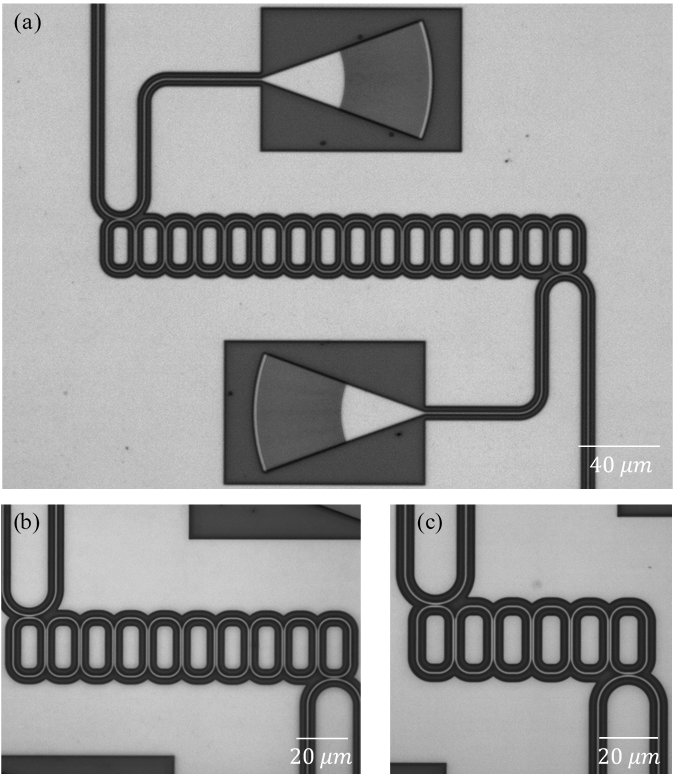}
	\caption{\label{fig:fig1001} Optical micrograph images of some of the fabricated photonic baths. (a) SSH photonic bath with 16 sites configured in trivial phase, fitted with grating couplers at each end. (b) Bath with 10 sites, (c) Bath with 6 sites.}
\end{figure*}

\clearpage

\begin{figure*}[t]
	\includegraphics[width=1\linewidth]{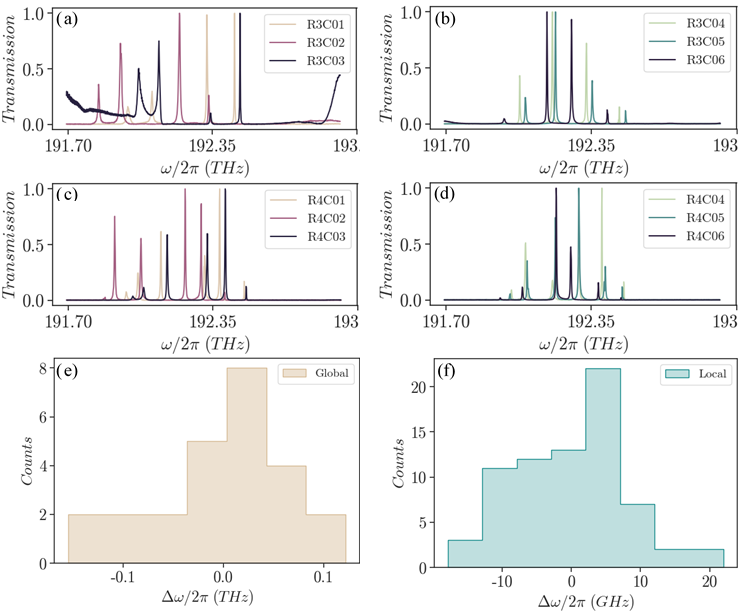}
	\caption{\label{fig:fig1002} Transmission spectra of some selected baths and disorder plots depicting deviations. Baths with $4$ resonator sites in (a) trivial phase, (b) topological phase. Baths with $6$ resonator sites in (c) trivial phase, (d) topological phase. Compiled results from baths with different sizes are used to calculate the disorder in system. (e) Histogram denoting deviations of mean frequency used to characterize the global disorder. (f) Histogram denoting deviations of mode frequencies from their respective mean positions after shifting to a common origin, used to characterize the local disorder.}
\end{figure*}

\clearpage

\section{Additional properties of the SSH photonic bath}
\begin{figure}[h]
	\includegraphics[width=1\linewidth]{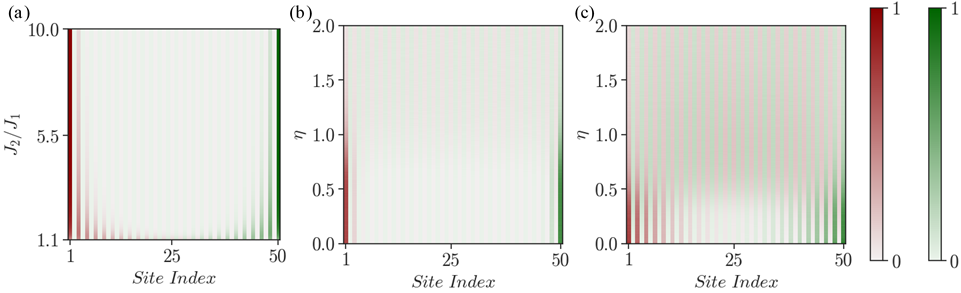}
	\caption{\label{fig:fig1003} Disorder study for edge states. Modulus of amplitude of wavefunctions for: (a) Hybridized edge states as we sweep across $J_2/J_1$ keeping $J_2-J_1$ constant. (b) Hybridized edge state amplitude averaged across ${10}^4$ disorder realizations per $\eta$ with $J_2/J_1\ =\ 6.25$. (C) Hybridized edge state amplitude averaged across ${10}^4$ disorder realizations per $\eta$ with $J_2/J_1\ =\ 1.33$. Sub-lattice A in red and sub-lattice B in green.}
\end{figure}

\begin{figure}[h]
	\includegraphics[width=0.5\linewidth]{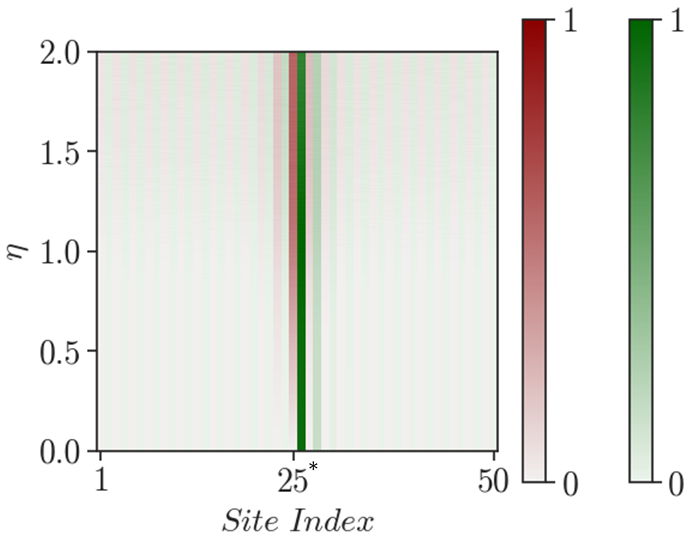}
	\caption{\label{fig:fig1004} Disorder study for photonic bound state when $J_1/J_2\ =\ 6.25$. Modulus of amplitude of photonic bound state averaged across ${10}^4$ disorder realizations per $\eta$. The zero detuned emitter is coupled to trivial bath at site $25$. Sub-lattice $A$ in red and sub-lattice $B$ in green.}
\end{figure}

The edge state localization length $\xi$ of a SSH bath in topological phase depends on the ratio of the hopping rates $J_2/J_1$  as \cite{asboth2016short}
\begin{align}\label{localization}
\xi=\frac{1}{log(\frac{J_2}{J_1})}
\end{align}
where $J_2>J_1$ for the topological phase. In Fig. \ref{fig:fig1003}a we plot the effect of this ratio on the edge states keeping $J_2-J_1$ constant (this ensures that band gap = $2|J_1-J_2|$ does not change). 
The wave function of the edge states is localized on the either side on respective sub-lattices, with an exponential tail into the bulk. As $J_2/J_1$ increases the localization length decreases and when $J_2/J_1>>1$, the edge modes are almost entirely localized on the first and last sites. In Fig. \ref{fig:fig1003}b, c we plot the effect of disorder on the edge states for two cases $J_2/J_1=6.25$ and $J_2/J_1=1.33$. As seen, a large $J_2/J_1$ causes the amplitude on each successive site to decay very rapidly. This approach was adopted in an older work to demonstrate topological edge states \cite{mittal2019photonic} despite a large disorder to hopping rate ratio $\eta$. But as discussed in the paper such a regime of operation though useful for observation of edge states, is not suitable for quantum simulation purposes. For using a photonic bath to mediate long range topological interactions of quantum emitters $\eta\ <1$ is required. In Fig. \ref{fig:fig1004} we demonstrate the effect of disorder for a bound state arising when an emitter with zero detuning is coupled to trivial SSH bath $(J_1>J_2)$ with $J_1/J_2$=6.25. As evident the bound state loses it special properties like chirality and sub-lattice localization when $\eta\ \rightarrow1$.  This plot is a counterpart to Fig. 2 in the main text where we plotted a similar figure but with $J_1/J_2=1.33$. A longer localization length allows the bath to mediate interactions over longer range which was the motivation for keeping a low $J_1/J_2$ ratio in our work. 
Additionally, by using the scheme proposed in our paper we get a free control over $J_2/J_1$ and consequently the localization length. This allows for realization of topological quantum photonic baths with varying physical range of interactions, as long as we operate in $\eta<<\ 1$ regime.

\clearpage

\begin{figure*}[t]
	\includegraphics[width=1\linewidth]{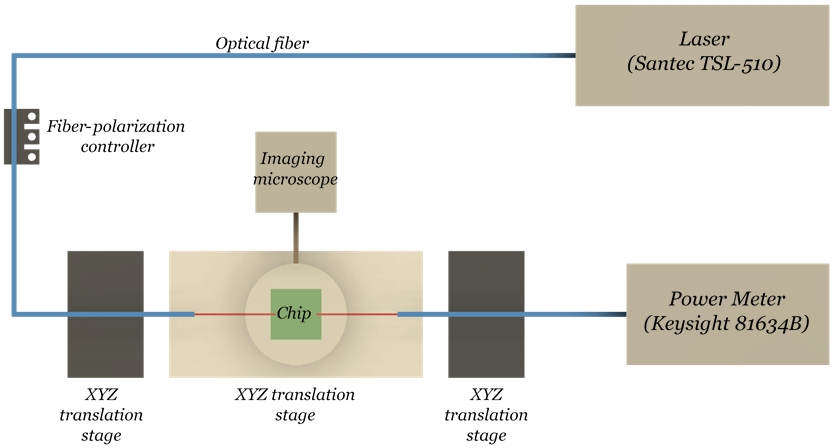}
	\caption{\label{fig:fig1005} Experimental setup for measuring transmission spectrum. A tunable continuous-wave laser (Santec TSL-510) is coupled to fiber and its polarization is controlled by a manual fiber-polarization controller (Thorlabs FPC526). The fiber is then coupled to the input grating of the device using a XYZ translation stage. The output grating is simultaneously coupled to a low noise power meter (Keysight 81634B) via similar fiber path. In order to aid alignment, the chip is continuously monitored via an imaging microspore (AmScope).}
\end{figure*}

\clearpage